# A Survey on Mobile Data Gathering in Wireless Sensor Networks - Bounded Relay

Ms.Rubia.R[1], Mr.SivanArulSelvan[2]

1 – Student, Department of CSE, Kumaraguru College of Technology, Coimbatore-49
2 – Associate Professor, Department of CSE, Kumaraguru College of Technology, Coimbatore-49
Tamil Nadu, India.

*Abstract*— Most of the wireless sensor networks consist of static sensors, which can be deployed in a wide environment for monitoring applications. While transmitting the data from source to static sink, the amount of energy consumption of the sensor node is high. It results in reduced lifetime of the network. Some of the WSN architectures have been proposed based on Mobile Elements. There is large number of approaches to resolve the above problem. It is found those two approaches, namely Single Hop Data Gathering problem (SHDGP) and mobile Data Gathering, which is used to increase the lifetime of the network. Single Hop Data Gathering Problem is used to achieve the uniform energy consumption. The mobile Data Gathering algorithm is used to find the minimal set of points in the sensor network, which serves as data gathering points for mobile network. Even after so many decades of research, there are some unresolved problems like non uniform energy consumption, increased latency, which needs to be resolved.

## I. INTRODUCTION

Wireless Sensor Networks (WSNs) consist of sensor nodes that are deployed into a large scale sensing field without a preconfigured infrastructure. The goal of the sensor node is to collect the data at regular intervals, then transform the data into digital signal and finally send the signal to the sink or the base node. Before monitoring the environment, the sensor nodes must identify their neighbour nodes and forms a network. Energy consumption can be takes place while sensing the field and uploading the data to Mobile Collector. The sensor networks can be classified into two types namely, homogeneous and heterogeneous networks.

The nodes in these networks having identical capabilities and energy in a network is called homogeneous network [2]. These types of networks can be again classified into flat and hierarchy topology. In the flat topology that the sensors close to the static sink consumes more energy than the sensors at the margin of the network.

In some applications, the sensor nodes are deployed to monitor different areas. In such applications, the network may be disconnected. In those applications, the sensors cannot forward data to sink via wireless links. A mobile collector can be used to collect the data. Mobile collector is a device equipped with powerful transceiver and high battery power [1].

The drawbacks of flat topology can be overcome by using hierarchical topology i.e., clusters. In this, the group of nodes that forms the lower layer and the cluster heads at the higher layer [2] [3]. Cluster head, which collects data from the lower layers and then forwards it to the sink. Cluster head can acts as an aggregation point. Since the cluster head is collecting data from lower nodes, it consumes more energy than other nodes. So, the sensor nodes can be rotated dynamically to avoid the energy consumption.

The Heterogeneous networks having small number of resource rich nodes and large number of resource limited basic nodes. The resource rich nodes are having powerful transceivers and batteries. The resource rich nodes can acts as cluster heads. The resource limited basic nodes having limited communication capabilities.

Mobile Data Gathering is a technique that consists of one or more Mobile Collectors (MC's) [1]. Mobile collector is a device equipped with powerful transceiver and high battery power. It gathers the data in short range communications. MC roams over the sensing field to collects the data while moving or pause at some points on its moving path from the sensors. To attain the maximum energy saving, a mobile collector must travel the transmission range of each sensor node in the field. It helps the mobile collector to collect the data packets in a single hop. The path of the mobile collector in the sensing field may be random or planned. The mobility of the collector reduces the energy consumption in the network.

Every sensor communicates directly with the sink is called single-hop relay [7]. It requires large transmit power and may be infeasible in large geographic areas. Sensors that serve as relay for other sensor nodes are known as multi-hop routing in wireless sensor networks. Data packets are forwarded to data sink via multi-hop relay among sensors. Energy consumption is more while forwarding the data packets in multi-hop.

To achieve the uniform energy consumption, the Single Hop Data Gathering Problem (SHDGP) is used [2]. The mobile Data Gathering algorithm is used to find the minimal set of points in the sensor network. It serves as data gathering points for mobile node [6] [8].

## II. DATA COLLECTION TECHNIQUES

The data collection technique is used to collect the aggregate data from the sensor node to the sink node. The main objective of the data collection process is to reduce the delay and improves the network's lifetime. There are various techniques used to collect the data from source node to sink node.

First, all the sensors are static and then the network is considered as static network. The static sensor node forwards the data to the sink by one or more hops [3]. So, the sensor located nearer to the sink gets depleted soon.

Second, the hierarchy form of data collection. The nodes can be categorized into lower layer and higher layer. The nodes in the lower level layers are homogenous sensor nodes. The nodes in the higher layer are more powerful than the nodes in the lower layer. The higher layer nodes are called as cluster heads. The hierarchy topology is also called as clusters.

Third, Mobile Collector is used to collect the data periodically. A mobile data observer is used to collect the data dynamically. The nodes that can be located closer to the data observer can upload the data directly. The nodes that can be located far away from the observer can forward the data by relaying [3].

Single Hop Data Gathering problem (SHDGP) and mobile Data Gathering are the two approaches that can be used to increase the lifetime of the network. Single Hop Data Gathering Problem (SHDGP) is used to achieve the uniform energy consumption. The mobile Data Gathering algorithm is used to find the minimal set of points in the sensor network. It serves as data gathering points for mobile node.

*A. Single Hop Data Gathering Problem (SHDGP)*

A Mobile data Collector can be represented as M-Collector. M-Collector is a device equipped with powerful transceiver and high battery power. It collects the data directly from the sensor node while it roams in the sensing field. By reducing the tour length of the M- collector, the lifetime of the sensor network can be prolonged. The M-collector visits the data in the transmission range of each sensor, in order to find the shortest moving tour.

The sensor nodes represent the polling points or the nodes in one-hop range of M-collector (see Fig. 1). By assuming that the M-collector moves at fixed speed, then the time consumption of the M-collector can be roughly estimated by using the tour length. If the M-collector travels in the shortest path, it results the data collection in shortest time. Thus, the users can collect the up-to-date data. This problem is referred as the single hop data gathering problem, or SHDGP.
The location of every sensor node can visited one by one by using the M-collector. The problem is reduced to Traveling Salesman Problem (TSP) [5]. The main objective of the TSP is to find the shortest distance (cost) tour that visits every node in the network atleast once. The sensors with fixed transmission power are deployed in large area that can be used in applications such as battlefield surveillance and environment monitoring.

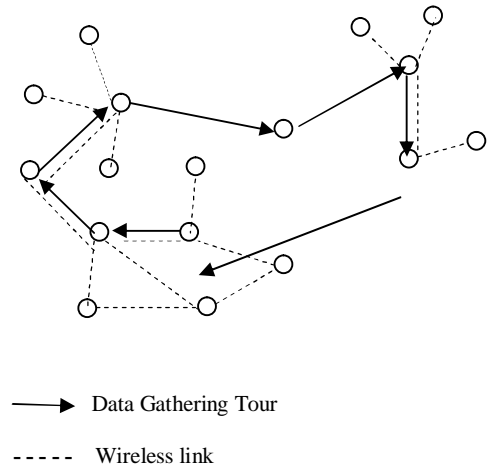

Fig. 1 Single Hop Data Gathering Problem

The sensor nodes that poll the data directly to the M-collector in single hop are called as *Polling Points* [1] [2]. The polling point transmits all the data in its transmission range. M-collector after collects the polled data from the polling point, it moves to the next polling point. The M-collector must traverse all the polling points in the network and finally reaches the static data sink.

For example, consider a set of polling points as P = {$p_1$, $p_2$, $p_3$ … $p_n$} and static data sink as S. The tour length of the M-collector can be denoted as S $\rightarrow$ $p_1$ $\rightarrow$ $p_2$ $\rightarrow$ $p_3$ $\rightarrow$ …. $\rightarrow$ $p_n$ $\rightarrow$ S. The problem is to find the optimal tour of the M-collector and how to find the polling points and order to visit those polling points.

M-collector needs to identify the polling points as well as its locations before starts its data gathering tour. The *neighbour set* of a point defined as the set of sensors. It can upload the data directly to the M-collector [2]. Each sensor in the neighbour set must have at least one polling point to upload the data in single hop. All the sensor nodes should be covered while combining the neighbour sets of all the polling points in the network.

It is impossible to find the neighbour set of an unknown point unless the M-collector traverse towards the polling point and test the wireless link between the M-collector and the sensor node or place the sensor at particular location and identify its one-hop neighbour while discovering the neighbour phase.

It is possible to test the finite number of points and its corresponding neighbour sets and select the polling points as *candidate polling points* [2]. If the one-hop neighbour of each sensor is known, the position of the sensor can be a candidate polling point.

A set of sensors, a set of candidate polling point, the starting point and ending point of the M-collector tour and the neighbour set of candidate polling point should be known to identify the polling points and determine the sequence of visits among the polling points. The above technique is used to minimize the total distance of data gathering tour of the M-collector.

*B. Mobile Data Gathering*

The main objective of the mobile Data Gathering technique is to reduce the overall travel time of the mobile node and also reduce the packet delay. Mobile Collectors are called as SenCars [6]. To achieve the uniform energy consumption, we combine the SDMA technique along with SenCar. This technique adopts a joint approach of Space Division Multiple Access and mobility [4]. The SDMA technique contains multiple antennas that help for concurrent data uploading to a SenCar. There are two cases namely single SenCar and multiple SenCar [8].

For a single SenCar, the main objective is to reduce the total data gathering time. It includes the travelling time of the SenCar and the uploading time of sensors to the SenCar. This problem is referred as mobile data gathering with SDMA (MDG-SDMA).

For multi-SenCar, the sensing field is divided into several areas. Each area is having a SenCar [8]. It mainly focuses on balancing the data gathering time on different regions. This problem is referred as mobile data gathering with multiple SenCars and SDMA (MDG-MS).

The mobility means deploying two or more SenCars in a sensing field that collects data from various sensors at particular location via single-hop transmissions. There are three advantages for the usage of mobile elements in the sensing field.

First, the non-uniform energy consumption can be reduced among the sensors. The sensor can upload the data directly to the SenCar rather than forwarding the data in multi-hop transmission. Second, it is suitable for connected network as well as disconnected network. The path of the SenCar can be considered as virtual links among separated sub networks [8]. Third, the tour of the SenCar can be predictable. It is useful for obtaining the optimal tour length of the SenCar.

*1) Mobile Data Gathering with a Single SenCar and SDMA technique (MDG-SDMA):*

A SenCar is equipped with two antennas and all the sensor nodes are equipped with single antenna are deployed in the sensing field. The sensor nodes that poll the data directly to the SenCar in single hop are called as *Polling Points* [8]. *Coverage area* is defined as the disk shaped area centred at the polling point with the radius equal to the sensor transmission range.

The *Neighbour set* is formed by the sensors in the coverage area of the polling point. Even though the sensors may locate at coverage area of multiple polling points, each sensor node needs to be polled only once during a data gathering tour, it is associated with only one polling point. If two sensor nodes are compatible, the compatible pair to be scheduled to upload the data simultaneously. A SenCar need not to be visited all the polling points in the sensing field. The polling points on the tour must cover the entire sensors in the sensing field. These polling points are called as *selected polling points*.

The SenCar arrives the selected polling points and collects data from all the associated sensors. Then moves to the next selected polling point and so on. The moving tour of the SenCar consists of number of selected polling points. The selected polling points are connected by using straight lines. For example, consider a set of selected polling points as P = $\{p_1, p_2, p_3 \ldots p_n\}$ and static data sink as DS. The tour length of the SenCar can be denoted as DS $\rightarrow p_1 \rightarrow p_2 \rightarrow p_3 \rightarrow \ldots \rightarrow p_n \rightarrow$ DS. The problem is to find the optimal tour of the SenCar and how to find the polling points and order to visit those polling points.

A series of problems needs to be solved. First, the SenCar must be able to determine whether the two sensors are compatible or not. Second, the SenCar must collect the data as fast as possible. It should identify the maximum number of compatible pairs. This can be formalized using the matching problem in a compatibility graph [8]. The vertex represents sensor and two vertices are adjacent to each other, then the sensors are said to be compatible. In graph theory, Matching is defined as a set of vertex-disjoint edges in the graph corresponds to a group of compatible pairs. The SenCar can collect the data in the place that has more compatible sensors. Hence, the data can be collected in shorter time.

To minimize the time of data uploading, the SDMA technique is used. To prolong the moving tour, the SenCar may have to visit some specific locations [8]. Consider the set of polling points as ℙ. The subset of ℙ can be denoted as ℙ'. By visiting the ℙ', all data can be collected in minimum time. The polling points in ℙ' are known as selected polling points.

*2) Mobile Data Gathering with Multiple SenCars and SDMA technique (MDG-MS):*

The single SenCar takes a long data gathering tour to collect the data. To avoid this problem, multiple numbers of SenCars can be deployed with SDMA technique to collect the data in the subareas.

In case of MDG-MS, the sensing field is divided into number of non-overlapping subfields. Each subfield is having a SenCar. Each SenCar can forward the collected data to another SenCar and so on. Finally, the data reaches the static data sink. The SenCar forwards the data once it collects all the data in the region or also forwards while they are moving on the paths except at the time of SenCars are communicating with its associated sensors.

Two sensors in the compatible pair would upload the data to the SenCar simultaneously. If the sensor is isolated, it would upload the data to the SenCar separately. The Sensor goes to sleep mode once it completes the process of data gathering in its region. It results the optimal data gathering tour by achieving the network's lifetime and minimizing the data gathering latency [8]. This problem is referred as Mobile Data Gathering with Multiple SenCars and SDMA technique. To balance the data gathering time among the different regions, the selected polling points and their associated sensors should be properly partitioned. The Region- Division and Tour Planning algorithm is used to find the short data gathering time by considering the whole sensing field in the single SenCar [8]. By considering the weight of the polling point and divide them into different regions based on the weight.

TABLE I
COMPARISON AMONG TWO MOBILE DATA GATHERING SCHEMES

|  | Mobile Data Gathering schemes | |
|---|---|---|
|  | SHDGP | MDG |
| Motion Pattern | Controllable, free to go anywhere | Controllable, with fixed moving tracks |
| Pausing Locations | Mobile Collector pauses at the location of PPs that can be chosen from a set of candidate PPs and collect the data from it | Exact pausing locations are not specified explicitly. MC always collects the data while moving along the tracks |
| Moving Trajectory | Start from the data sink; visit some locations that cover the transmission range of mobile collector. Finally it reaches the data sink | Start from the data sink, go along the parallel straight tracks back and forth, and finally go back to the data sink |
| Relay for Local Data Aggregation | No local relays | Multi-hop relays |
| Data uploading | Each sensor directly uploads data to the mobile collector in a single hop when it arrives within its range | Some sensors close to the tracks upload aggregated packets to the mobile collector when it comes |

The table 1 shows the comparison between the two data gathering schemes by comparing the motion pattern, pausing locations, moving trajectory, relay for local data aggregation and data uploading.

## III. ISSUES

There are some other issues in Wireless Sensor Networks such as load balancing [9], schedule pattern [10] and data redundancy [11]. If the energy consumption is more, then the lifetime of the network will be less. While transmitting the data from the sensor node to sink node, the latency will be high [12]. There is unnecessary energy consumption in the case of multi-hop routing. Mobile Collector cannot cover all the sensor nodes if it moves along the straight line.

Some of the design challenges of WSN are power consumption and production cost, reliability, scalability, mobility, bandwidth and responsiveness. The power consumption and production cost of the sensor nodes will be high. There is a limited computational power and memory size of each sensor node. As the number of node increases, the overhead of the network gets increases.

By achieving the high responsibility, the reliability and scalability can be achieved automatically .By enhancing the mobility concepts, the lifetime of the network can be greatly increased [13]. Limited bandwidth results in congestion that affects the normal data exchange.

## IV. CONCLUSIONS

The above data gathering schemes for large scale networks are discussed. A mobile data collector is introduced, like mobile base station. The Single-Hop data gathering scheme improves the scalability. It also solves the intrinsic problems large homogenous networks. The above scheme is suitable only for the partially connected applications. In large scale applications, there are strict time/distance constraints. For mobile data gathering, the mobility and SDMA are jointly considered. A single SenCar is used to improve the data gathering tour. Multiple SenCars are used to shorten the data gathering tour.